\documentclass[times, 10pt,twocolumn]{article} 

\usepackage{amsthm}
\usepackage{amsmath}
\usepackage{amssymb}
\usepackage{times}
\usepackage[dvips]{graphicx}

\newtheorem{theorem}{Theorem}
\newtheorem{example}{Example}

\begin{document}

\title{Degeneracy of Angular Voronoi Diagram}

\author{
	Hidetoshi Muta\\
	Department of Computer Science, \\
	University of Tokyo \\
	hmuta@is.s.u-tokyo.ac.jp
	\and
	Kimikazu Kato\\
	Nihon Unisys, Ltd. \\
	\& \\
	Department of Computer Science, \\
	University of Tokyo \\
	kkato@is.s.u-tokyo.ac.jp
}
\date{}

\maketitle

\begin{abstract}
Angular Voronoi diagram was introduced by Asano et al.\ as fundamental research for a mesh generation. 
In an angular Voronoi diagram, the edges are curves of degree three. 
From view of computational robustness we need to treat the curves carefully, because they might have a singularity. 

We enumerate all the possible types of curves that appear as an edge of an angular Voronoi diagram, 
which tells us what kind of degeneracy is possible and 
tells us necessity of considering a singularity for computational robustness.
\end{abstract}


\section{Introduction}

The Voronoi diagram has developed as one of the most important research objects in computational geometry~\cite{vdsurvey, spatialTessellations}.
Although originally Voronoi diagrams were investigated in the terms of Euclidean space, 
research about Voronoi diagrams in other distance spaces has developed recently. 
Voronoi diagrams can be considered as a powerful tool to analyze the structure of an unnatural distorted distance space~\cite{kato05,kato06a,kato06b}.

Asano et al.~\cite{avd} first introduced the angular Voronoi diagram. 
It is proposed as a fundamental research applicable to a mesh generation. 
They showed edges of an angular Voronoi diagram are curves whose degree is at most three. 

\begin{figure}
	\begin{center}
	\includegraphics[width=11em,clip]{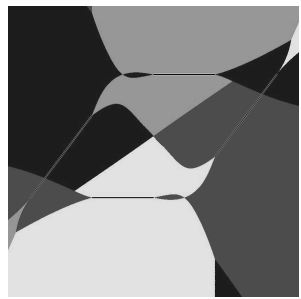}
	\end{center}
	\caption{Examples of where two edges are crossing at their
 singular points} \label{sample}
\end{figure}

Generally understanding degeneracy of a Voronoi diagram is very important especially from a computational point of view. 
It is a basis to assure the robustness of computation of the actual diagram. 
To deal with other distance spaces and keep computational robustness, we extend the meaning of ``degeneracy''.
When we say a ``degenerate case,'' it is a case which takes special care for computational robustness. 
Especially for an angular Voronoi diagram we have to consider singularities of curves. 
We show a symbolic example in Fig.~\ref{sample}. 
In this example, the two edges have a intersection at their singular points, both of which are nodes.

In this paper, we show a classification of possible curves which can appear as edges of an angular Voronoi diagram. 
Although we have not completed the whole classification, we have achieved to show possible variety of curves. 
This is a first step to investigate the degeneracy angular Voronoi diagrams. 
This is a good indication for a structure of a general Voronoi diagram with respect to non-Euclidean distance.

The rest of this paper is organized as follows.
First, in Sect.~\ref{preliminary}, we give some mathematical properties of degree three equation and properties of edges of an angular Voronoi diagram.
In Sect.~\ref{threeDegreeCase} and Sect.~\ref{lowerDegreeCase} , we show degeneracy of angular Voronoi diagram.
We give a conclusion in Sect.~\ref{conclusion}.




\section{Preliminaries}\label{preliminary}


\subsection{Curves of degree three and their singularities}

We explain basic mathematical facts about an algebraic curve
of degree three. For more general theory about algebraic curves, see
\cite{Hartshorn77}. As for singularities, three has a special meaning as
a number of degree of curve, because curves of degree two have only
a trivial singularity. Actually, a curve of degree two is singular
only when its equation is divided into two linear equations, and the only
singularity that may appear is the cross point of the lines.

Generally for a curve $ F(x, y) = 0 $, if $ \frac{\partial F}{\partial x} = 0 $
 and $ \frac{\partial F}{\partial y} = 0 $, the point is called a singular point.
Otherwise, the point is called a regular point. 



\begin{figure}
	\begin{center}
	\includegraphics[width=7.5em,clip]{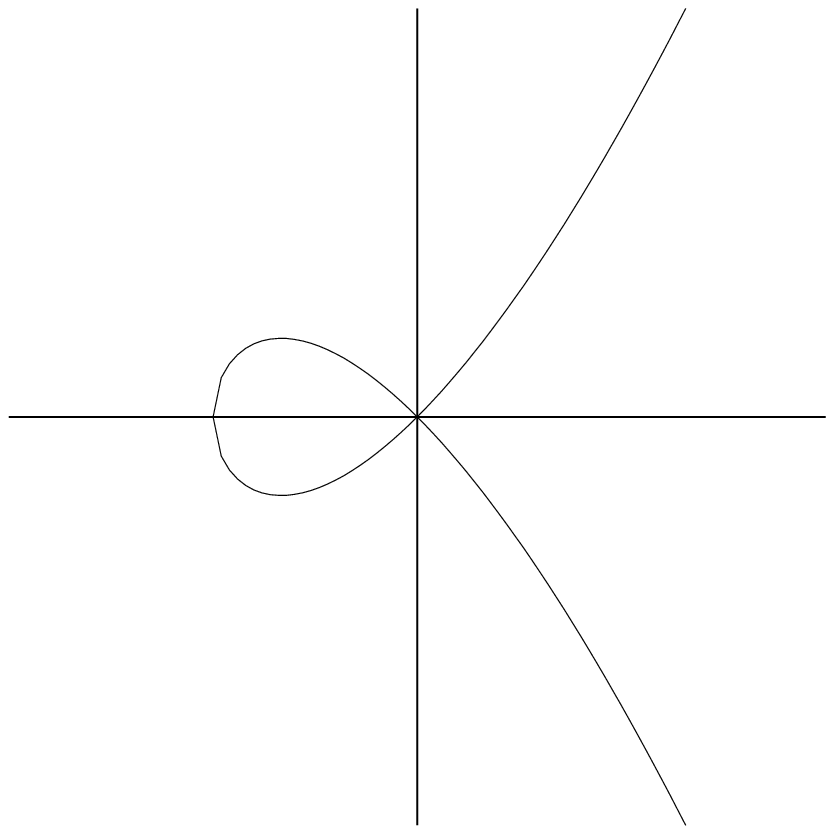}
	\includegraphics[width=7.5em,clip]{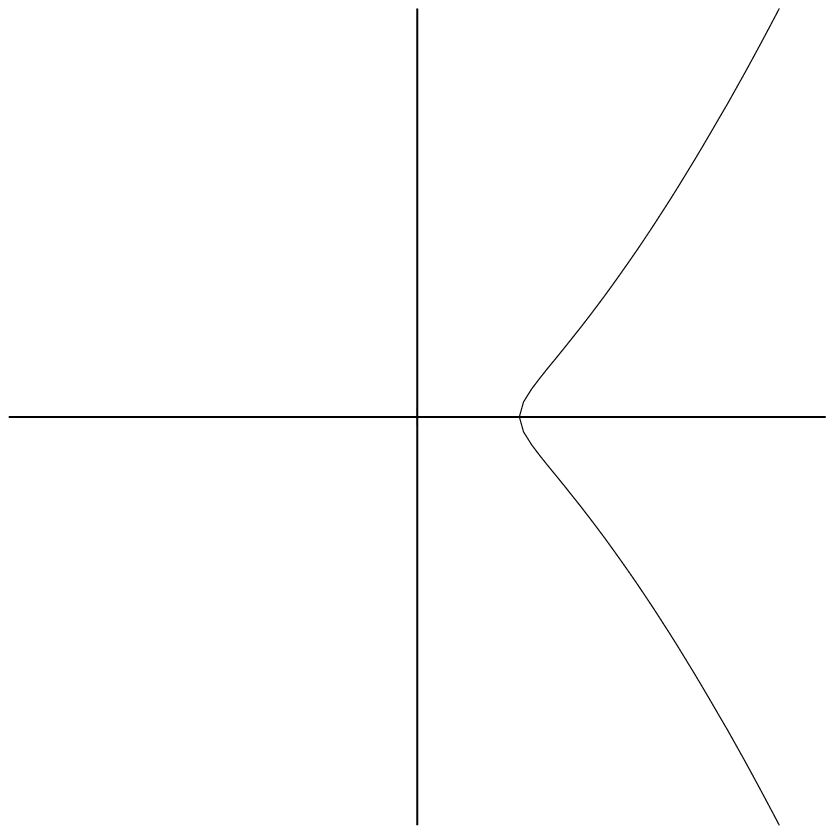}
	\includegraphics[width=7.5em,clip]{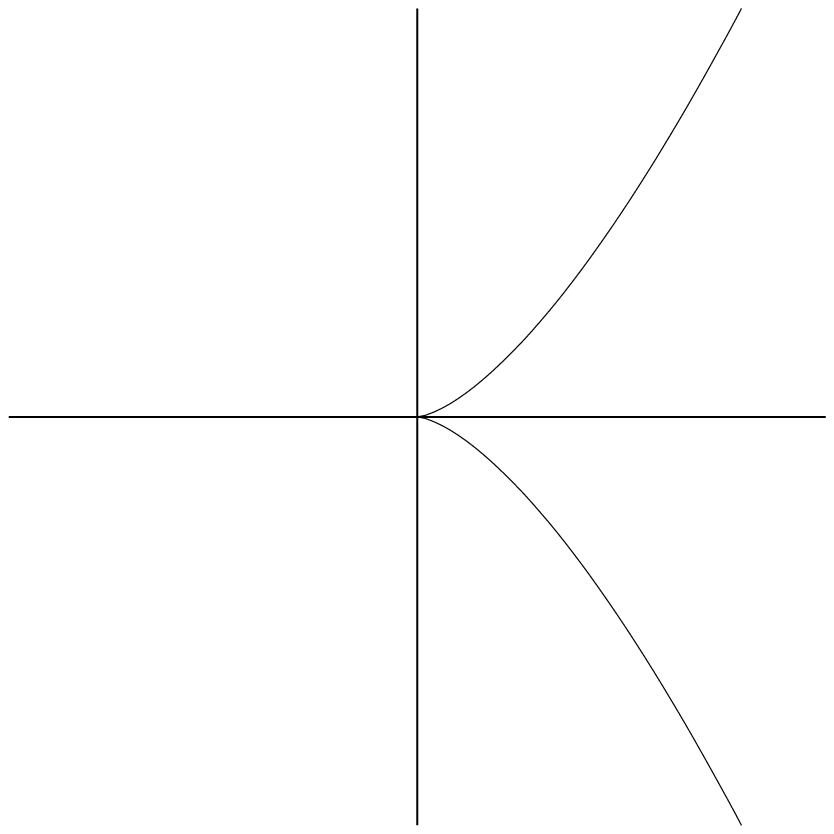}\\
	$ ( 1 ) $
	\qquad\qquad\qquad$ ( 2 ) $
	\qquad\qquad\qquad$ ( 3 ) $
	\end{center}
	\caption{Three types of singular points of curves of degree three}
	\label{sng}
\end{figure}

We show three examples of singular points. Let $ F(x, y) = y^2 - x^2 (x + a) = 0 $, that is $ y = \pm x \sqrt{x + a} $. 
$ \frac{\partial F}{\partial x} = - 3 x^2 - 2 a x$, $ \frac{\partial F}{\partial y} = 2 y $, 
then the point $ (0, 0) $ is a singular point, 
Additionally, the singular point is called as one of the following three types as shown in Fig.~\ref{sng}.

$ ( 1 ) $ $ a > 0 $. 
Two smooth curves, which are symmetric with respect to the $x$-axis, pass through the point (0, 0) as shown in the left of Fig~\ref{sng}, 
because $ x + a > 0 $ in the neighborhood of $ x = 0 $.
In this case, $ (0, 0) $ is called a node.

$ ( 2 ) $ $ a < 0 $. 
There is no real corresponding $ y $ value as shown in the middle of the Fig~\ref{sng}, since $ x + a < 0 $ in the neighborhood of $ x = 0 $, 
$ (0, 0) $ is an isolated point.

$ ( 3 ) $ $ a = 0 $. Then $ y = \pm x^{\frac{3}{2}} $. If $ x > 0 $, 
there are two curves which  are symmetric with respect to the $x$-axis as shown in the right of the Fig~\ref{sng}.
The two curves have the common tangential line on $ (0, 0) $, This singular point is called a cusp.


\subsection{Angular Voronoi diagram}
\label{avdedge}
In this section, we define the angular Voronoi diagram and show its edge is at most three degree.
When a line segment $ s $ and a point $ p $ in a plane are given, 
the visual angle of $ s $ from $ p $ is the angle formed by two rays emanating from $ p $ through two endpoints of $ s $, 
and it is denoted by $ \theta_p(s) $. Since we do not use the direction of the angle, the range of the angle is always between $ 0 $ and $ \pi $.


Given a set $ S $ of $ n $ line segments $ s_1, s_2, \dots, s_n $ in a plane, 
we define an angular Voronoi diagram $ AVD(S) $ for $ S $ as follows:

\begin{description}
\item[Voronoi region:] 
Each line segment $ s_i $ is associated with a region, called a Voronoi region $ V(s_i) $, 
consisting of all points $ p $ such that the visual angle of $ s_ i $ from $ p $ is smaller 
than that of any other line segment $ s_j $.
\begin{equation}
	V(s_i) = \{ p \in \mathbb{R}^2 | \theta_p(s_i) < \theta_p(s_j) \text{ for any } j \neq i \}.
\end{equation}

\item[Voronoi edge:]
Voronoi edges form the boundary of Voronoi regions. Thus, they are defined for pairs of line segments:
\begin{equation}
	\begin{split}
	E(s_i, s_j) = \{ p \in \mathbb{R}^2 | \theta_p(s_i) = \theta_p(s_j) < \theta_p(s_k) \\ 
		\text{ for any } k \neq i , j\}.
	\end{split}
\end{equation}


\end{description}

The following theorem was proved by Asano et al.

\begin{theorem}[\cite{avd}]
	\label{asano}
	Edges of an angular Voronoi diagram are described by polynomial curves of degree at most three.
\end{theorem}

Let us consider a pair of line segments $ (s_1, s_2) $. 
Without lose of generality, we assume that $ s_1 $ is fixed on the $x$-axis, 
one of the endpoints of $ s_1 $ is $(1, 0)$ and the other is $ (-1, 0) $, 
and assume that the length of $ s_2 $ is $ 2 l $, 
the midpoint of $ s_2 $ is the coordinate $ ( a, b ) $ and
a slope between the $x$-axis and $s_2$ is $ \alpha $ (see Fig.~\ref{arrangement2seg}).

Then, we have the equation of the edge of the angular Voronoi diagram as follows:

\begin{figure}
	\begin{center}
	\includegraphics[width=20em,clip]{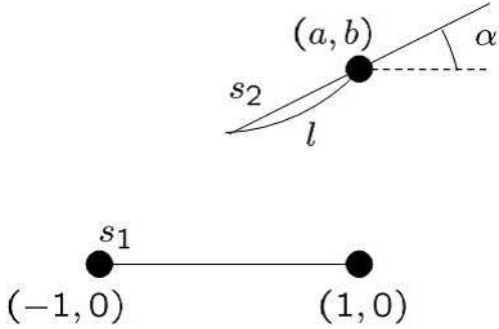}
	\end{center}
	\caption{The arrangement of the two line segments}
	\label{arrangement2seg}
\end{figure}

\begin{equation}
	\begin{split}
		& y \{ (x-a)^2 + (y-b)^2 - l^2 \} \\
		& -l \{ (x-a) \sin \alpha - (y-b) \cos \alpha \} ( x^2 + y^2 - 1 ) \\
		& = 0\label{eq:avdeq}
	\end{split}
\end{equation}

For the details of the proof of this theorem, see~\cite{avd}.

\section{Degree three case}\label{threeDegreeCase}
In this section, we show special cases even when the edge is of degree three. 
First, we show the condition for an edge to be a union of two curves. 
Second, we show the example of an edge which is an irreducible but has a singularity.
\begin{theorem}
\label{th:d3f}
	Whenever an equation of an edge of an angular Voronoi diagram is exactly degree three and factorable, 
	the equation is a product of a circle and a line.
\end{theorem}
\begin{proof}
	We expand Eq.~\eqref{eq:avdeq}, which is an equation of an edge of an angular Voronoi diagram. 
	Then, we have (we show only degree three terms of the equation, because it is too long.)
	\begin{equation}
		\begin{split}
			& ( l \cos \alpha + 1) y^3 + ( - l \sin \alpha ) y^2 x \\
			& + ( l \cos \alpha + 1) y x^2  + ( - l \sin \alpha) x^3 + \dotsb = 0. \label{eq:avdeqex} \\
		\end{split}
	\end{equation}
	The general form of a product of an expression of degree two and an expression of degree one is
	\begin{equation}
		\begin{split}
			& ( a_1 y^2 + a_2 y x + a_3 x^2 + a_4 y + a_5 x + a_6 )  \\
			& \times ( b_1 y + b_2 x + b_3)= 0. \label{eq:gened2d1} \\
		\end{split}
	\end{equation}
	We expand Eq.~\eqref{eq:gened2d1}. Then we have (only degree three terms)
	\begin{equation}
		\begin{split}
			& a_1 b_1 y^3 + (a_1 b_2 + a_2 b_1) y^2 x \\ 
			& + (a_2 b_2 + a_3 b_1) y x^2 + a_3 b_2 x^3 + \dotsb = 0. \label{eq:gened2d1ex} \\
		\end{split}
	\end{equation}
	If Eq.~\eqref{eq:avdeq} is factorable, 
	then both Eq.~\eqref{eq:avdeqex} and Eq.~\eqref{eq:gened2d1ex} have the same factors.
	\begin{align}
		 l \cos \alpha + 1 &= a_1 b_1, & -l \sin \alpha &= a_1 b_2 + a_2 b_1, \notag \\
		 l \cos \alpha + 1 &= a_2 b_2 + a_3 b_1, & -l \sin \alpha &= a_3 b_2. \label{simueq}
	\end{align}
	Eliminating $ l $ and $ \alpha $ from Eq.~\eqref{simueq}, we have
	\begin{align}
		 a_1 b_1 &= a_2 b_2 + a_3 b_1, & a_3 b_2 &= a_1 b_2 + a_2 b_1. \label{a5}
	\end{align}
	Solving Eq.~\eqref{a5}, we have
	\begin{gather}
		b_1 = b_2 = 0 \qquad  \text{or} \label{b1b20} \\
		a_1 = a_3, \qquad a_2 = 0. \label{a1eqa3a20}
	\end{gather}
	If Eq.~\eqref{b1b20} holds, then we have
	\begin{equation}
		b_3( a_1 y^2 + a_2 y x + a_3 x^2 + a_4 y + a_5 x + a_6 ) = 0 \label{eqd2} \\
	\end{equation}
	In this proof, we do not consider this case, because Eq.~\eqref{eqd2} is degree two.
	If Eq.~\eqref{a1eqa3a20} holds, then we have
	\begin{equation}
		( a_1 y^2 + a_1 x^2 + a_4 y + a_5 x + a_6 ) ( b_1 y + b_2 x + b_3)= 0 \\
	\end{equation}
	$ a_1 y^2 + a_1 x^2 + a_4 y + a_5 x + a_6 = 0 $ is the equation of a circle. 
\end{proof}
Note we do not say a radius of the circle is positive, that is to say, it may be 0 or imaginary number.

We found the following remarkable examples.
\begin{figure}
	\begin{center}
	\includegraphics[width=11em,clip]{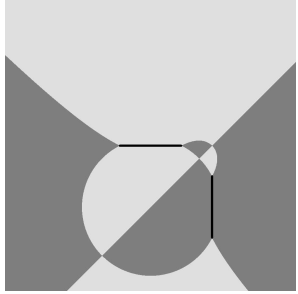}
	\end{center}
	\caption{Example 1 circular degeneracy}
	\label{circleex1}
\end{figure}
\begin{example}
	If four endpoints of two line segments are on the same circle and the two line segments have the same length, 
	an equation of an edge of an angular Voronoi diagram is a product of a circle and a line as in Fig.\ref{circleex1}.
\end{example}
\begin{proof}
	We use the same setting as Section \ref{avdedge}.
	We consider a circle $ C $ whose center is $ (0, h) $ and whose radius is $ \sqrt{h^2 + 1} $.
	Two endpoints of $ s_1 $ are on $ C $. 
	We change the parameters:
	\begin{align}
		a &= h \cos \theta, & b &= h + h \sin \theta, & l &= 1, &  \alpha &= \theta - \frac{\pi}{2}. \label{circlep1}
	\end{align}
	Then two endpoints of $ s_2 $ are on $ C $ and the length of $ s_2 $ is also $ 2 $.
	Substituting Eq.~\eqref{circlep1} into Eq.~\eqref{eq:avdeq} and factorizing it, then we have

	\begin{equation}
		\begin{split}
			& (y^2 + x^2 - 2hy - 1) \\
			& \times \{ ( 1 + \sin \theta) y + (\cos \theta) x - h ( 1 + \sin \theta) \} = 0 \label{circleeq2}.
		\end{split}
	\end{equation}
	Eq.~\eqref{circleeq2} is a product of an equation of a circle and an equation of a line. 
\end{proof}
\begin{figure}
	\begin{center}
	\includegraphics[width=11em,clip]{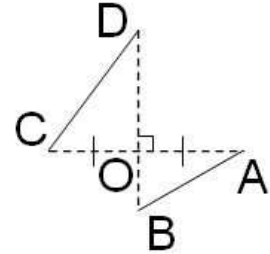}
	\end{center}
	\caption{Arrangement of two lines}
	\label{circlese}
\end{figure}
\begin{figure}
	\begin{center}
	\includegraphics[width=11em,clip]{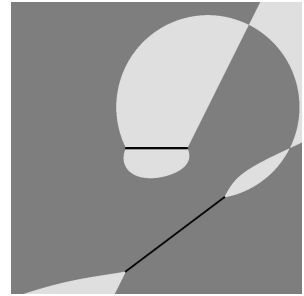}\\
	\end{center}
	\caption{Example 2 circular degeneracy}
	\label{circleex2}
\end{figure}
\begin{example}
	Suppose that two line segments AB and CD are given and O is a crosspoint AC and BD.
	If AC crosses BD orthogonally, 
	and if 
	the length of AO is equal to the length of CO and
	the length of BO is not equal to the length of DO in Fig.\ref{circlese}, 
	an equation of an edge of an angular Voronoi diagram is a product of a circle and a line as in Fig.\ref{circleex2}.
\end{example}
\begin{proof}
	We place line segment $ s_1 $ on $ (2, 0) , (0, - 2 \tan \theta_1) $ and
	place line segment $ s_2 $ on $ (-2, 0) , (0, 2 \tan \theta_2) $.
	Suppose $(x, y)$ is a point from where the visual angle of $ s_1 $ is equal to that of $ s_2 $.
	Then there are two circle $C_1$ and $C_2$, $C_1$ passes through $(x, y)$ and two endpoints of $s_1$ and $C_2$ passes similarly.
	We find the center of $C_1$ is $ (1 - h_1 \sin \theta_1, - \tan \theta_1 + h_1 \cos \theta_1) $, because $C_1$ passes two endpoints of $s_1$.
	In the same way, the center of $C_2$ is $ (-1 - h_2 \sin \theta_2, \tan \theta_2 + h_2 \cos \theta_2) $. 
%
%
%
%
	Because both $ (2, 0) $ and $ (x, y) $ are on $ C_1 $, 
	\begin{equation}
		\begin{split}
				& (1 - h_1 \sin \theta_1 - 2)^2 + ( - \tan \theta_1 + h_1 \cos \theta_1 )^2 \\
			=	& (1 - h_1 \sin \theta_1 - x)^2 + ( - \tan \theta_1 + h_1 \cos \theta_1 - y)^2. \label{p1}
		\end{split}
	\end{equation}
	Because both $ (-2, 0) $ and $ (x, y) $ are on $ C_2 $, 
	\begin{equation}
		\begin{split}
				& (-1 - h_2 \sin \theta_2 - (-2))^2 + ( \tan \theta_2 + h_2 \cos \theta_2 )^2 \\
			=	& (-1 - h_2 \sin \theta_2 - x)^2 + ( \tan \theta_2 + h_2 \cos \theta_2 - y)^2. \label{p2}
		\end{split}
	\end{equation}
	The ratio between the distance the center of $C_1$ to $s_1$ and the length of $s_1$ is equal to the ratio between the distance the center of $C_2$ to $s_2$ and the length of $s_2$, 
	since the visual angle of $s_1$ from $(x, y)$ is equal to that of $s_2$ from $(x, y)$.
	\begin{equation}
		\begin{split}
			h_1 : h_2 = \left| \frac{2}{\cos \theta_1} \right| : \left| \frac{2}{\cos \theta_2} \right|. \label{p3}
		\end{split}
	\end{equation}
	We expand Eq.~\eqref{p1} and Eq.~\eqref{p2} and substitute the results into Eq.~\eqref{p3}, then we have 
	\begin{equation}
		\begin{split}
			(x^3 + x y^2-4 x ) \sin(\theta_1 - \theta_2) - 4 x y \cos(\theta_1 - \theta_2) = 0. \label{p4}
		\end{split}
	\end{equation}
	We can divide Eq.~\eqref{p4} by $ \sin(\theta_1 - \theta_2) ( \neq 0 )$, because the length of BO is not equal to that of DO.
	\begin{equation}
		\begin{split}
			x \Big\{ x^2 + \left(y - 2\tan(\frac{\pi}{2} - \theta_1 + \theta_2)\right)^2 \\
			- \frac{1}{4 \cos^2(\frac{\pi}{2} - \theta_1 + \theta_2)} \Big\} = 0. \label{p5}
		\end{split}
	\end{equation}
	Eq.~\eqref{p5} is a product of an equation of a circle and an equation of a line. 
\end{proof}
\begin{figure}
	\begin{center}
	\includegraphics[width=11em,clip]{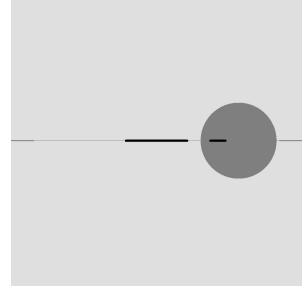}
	\end{center}
	\caption{Example 3 circular degeneracy}
	\label{circleex3}
\end{figure}
\begin{example}
	If two line segments are on the same line and their lengths are not equal, 
	an equation of an edge of an angular Voronoi diagram is a product of a circle and a line (see Fig.\ref{circleex3}).
\end{example}
\begin{proof}
	We use the same setting as Section \ref{avdedge}.
	$ s_1 $ is on the x-axis.
	Suppose $b, l$ and $ \alpha$ satisfy the following.
	\begin{align}
		b	& = 0,  & \alpha	& = 0, \pi,  & l	& \neq 1. \label{circlep2}
	\end{align}
	Then $ s_2 $ is on the $x$-axis and its length is not equal to the length of $ s_1 $.
	Substituting Eq.~\eqref{circlep2} into Eq.~\eqref{eq:avdeq} and factorizing, then we have
	\begin{equation}
		\begin{split}
			y \{ (\pm l + 1) y^2 + (\pm l + 1) x^2 - 2 a x + (a^2 + l^2 \mp l)\} = 0 \label{p6}.
		\end{split}
	\end{equation}
	Eq.~\eqref{p6} is a product of an equation of a circle and an equation of a line. 
\end{proof}
\begin{figure}
	\begin{center}
	\includegraphics[width=11em,clip]{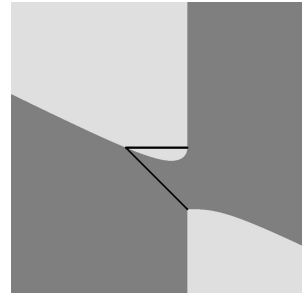}
	\end{center}
	\caption{Example 4 circular degeneracy}
	\label{circleex4}
\end{figure}
\begin{example}
	If two line segments share an endpoint, 
	an equation of an edge of an angular Voronoi diagram is a product of a circle and a line in Fig.\ref{circleex4}.
\end{example}
\begin{proof}
	We use the same setting of Section \ref{avdedge}.
	Suppose $a, b, l$ and $ \alpha$ satisfy the following.
	\begin{align}
		a	& = l \cos \alpha - 1, &	b	& = l \sin \alpha. \label{circlep3}
	\end{align}
	Then one of the endpoints of $ s_2 $ is also $ (-1, 0 ) $.
	Substituting Eq.~\eqref{circlep3} into Eq.~\eqref{eq:avdeq} and factorizing, then we have
	\begin{equation}
		\begin{split}
			\{y^2 + (x+1)^2 \} \{y(l \cos \theta - 1) - l (x -1) \sin \theta \} = 0
		\end{split}
	\end{equation}
	Eq.~\eqref{p5} is a product of an equation of a circle (in this case only a point) and an equation of a line. 
\end{proof}
We have discussed the case when the equation of the edge can be divided into two
polynomials. That is not the only ``unusual'' case. Even when an edge is also
an irreducible curve, there is an ``unusual'' case. The following is the
example.
\begin{example}
\label{th:d3u}
The following curve which appears as an edge of an angular Voronoi
 diagram is irreducible but has a node.
	\begin{align}
		a &= 2, & b &= \frac{4}{3}, & l &= \frac{5}{3}, & \cos \alpha &= \frac{3}{5}, & \sin \alpha &= - \frac{4}{5}. \label{nodepara}
	\end{align}
\end{example}
\begin{proof}
	Substituting Eq.~\eqref{nodepara} into Eq.~\eqref{eq:avdeq}, we have
	\begin{equation}
		\begin{split}
			f(x, y) = 2y^3 + \frac{4}{3} y^2 x + 2 y x^2 + \frac{4}{3} x^3 \\ 
			- \frac{20}{3} y^2 - 4 y x - 4 x^2 + 2 y - \frac{4}{3} x + 4. \label{nodeeq1}
		\end{split}
	\end{equation}
	We partially differentiate Eq.~\eqref{nodeeq1} as
	\begin{equation}
		\begin{split}
			\frac{\partial f}{\partial x} & = \frac{4}{3} y^2 + 4 y x + 4 x^2 - 4 y - 8 x - \frac{4}{3} \\
			\frac{\partial f}{\partial y} & = 6 y^2 + \frac{8}{3} y x + 2 x^2 - \frac{40}{3} y - 4 x + 2 \label{nodeeq3}.
		\end{split}
	\end{equation}
	Substituting $ y = 2 $ and $ x = - 1 $ into Eq.~\eqref{nodeeq1} and Eq.~\eqref{nodeeq3}, we have
	\begin{align}
		f = \frac{\partial f}{\partial x} = \frac{\partial f}{\partial y}  = 0.
	\end{align}
	The point (-1, 2) is a singularity and we find it is a node by drawing the graph of Eq.~\eqref{nodeeq1} (see Fig.\ref{fignode}).
	
	\begin{figure}
		\begin{center}
		\includegraphics[width=11em,clip]{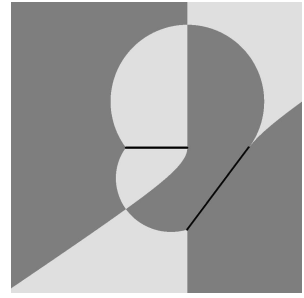}
		\end{center}
		\caption{Example of node}
		\label{fignode}
	\end{figure}
\end{proof}
\section{Lower degree case}
\label{lowerDegreeCase}
\begin{figure}
	\begin{center}
	\includegraphics[width=11em,clip]{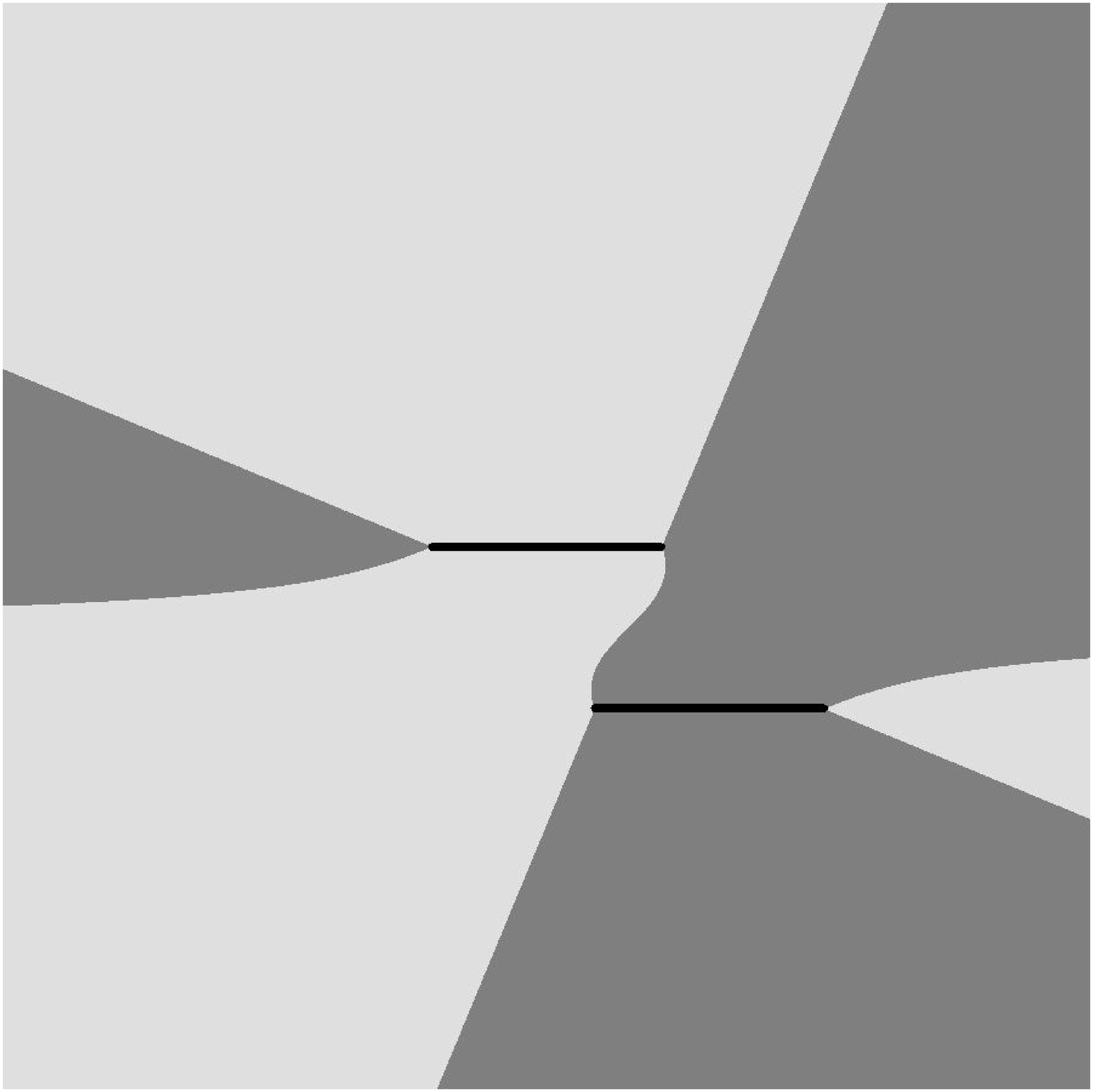}
	\includegraphics[width=11em,clip]{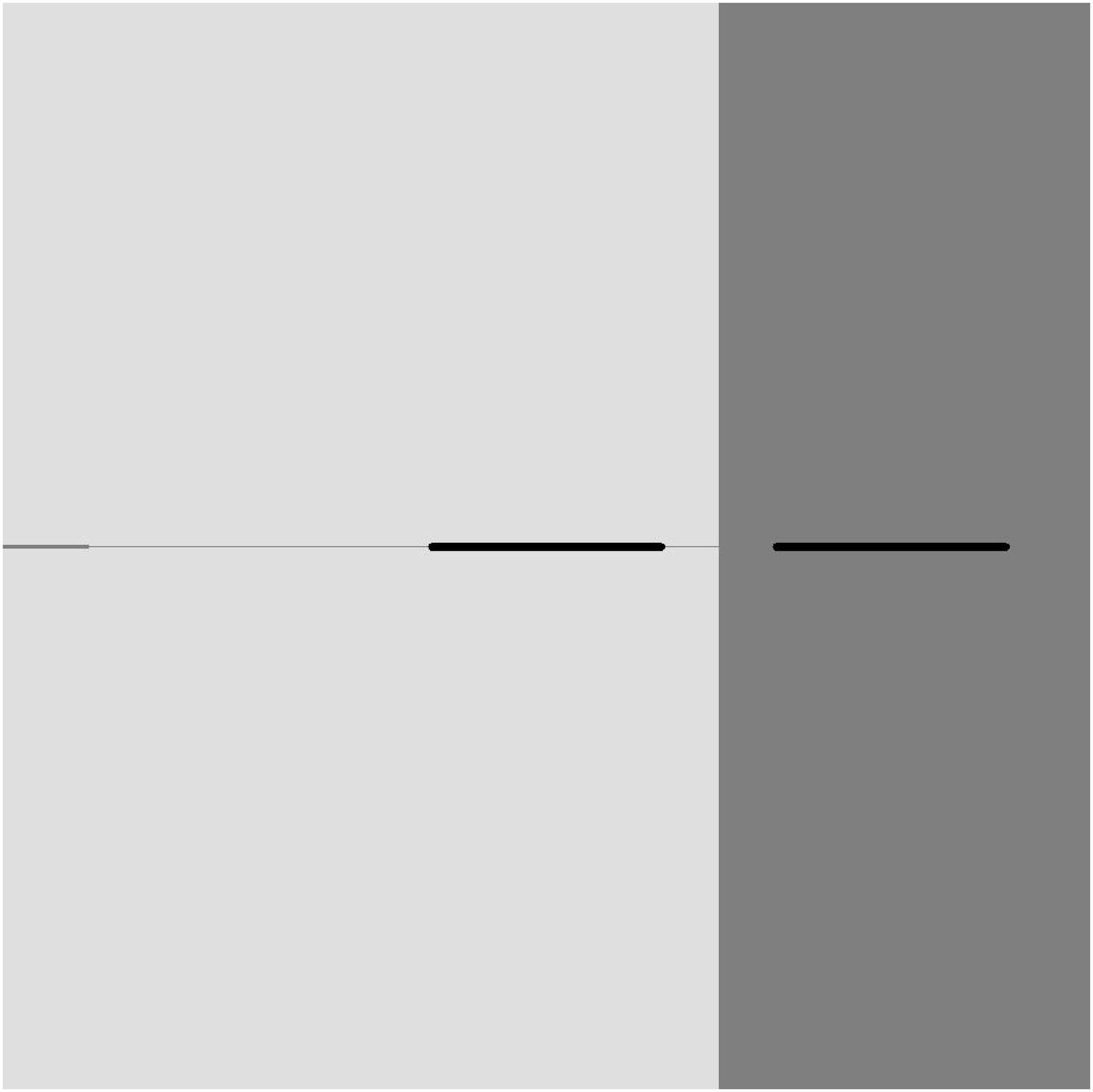}
	\end{center}
	\caption{Examples of Two orthogonal lines}
	\label{twoline}
\end{figure}

\begin{figure}
	\begin{center}
	\includegraphics[width=11em,clip]{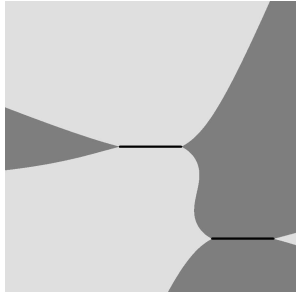}
	\end{center}
	\caption{Example of an orthogonal hyperbola}
	\label{hyperbola}
\end{figure}

In previous section, we explain degree three cases. 
Now, in this section, we explain degree less than three cases. 
A degree of an edge of an angular Voronoi diagram can be less than three. 
Three theorems in this section give a classification of all possible edges of degree less than three. 
In Th.~\ref{th:d2u} and Th.~\ref{th:d2f}, we show that the possible shapes for an edge of degree two are hyperbola or two lines.
Th.~\ref{th:d1} shows that a curve of degree one cannot appear as an edge.
\begin{theorem}
\label{th:d2u}
	An equation of an edge of angular Voronoi diagram is exactly degree two and irreducible, 
	if and only if the equation is an orthogonal hyperbola.
\end{theorem}
\begin{proof}
	The conditions for the degree three terms of Eq.~\eqref{eq:avdeqex} to be eliminated is
	$ l \cos \alpha = -1 $ and $ l \sin \alpha = 0 $.
	Solving this system, we derive
	\begin{align}
		l  & = 1, & \alpha = \pi. \label{p8}
	\end{align}
	Eq.~(\ref{p8}) means that the two line segments have the same length and are parallel.
	Substituting Eq.~\eqref{p8} into Eq.~\eqref{eq:avdeq}, we have
	\begin{equation}
		-b y^2 -2 a y x + b x^2 + (a^2 + b^2) y - b = 0. \label{d2eq}
	\end{equation}
	If $ b = 0 $, Eq.~\eqref{d2eq} is $ -2ayx + (a^2+b^2)y = 0 $.
	In this case the equation of the edge is $ y \{ 2 a x - (a^2+b^2) \} = 0 $. 
	It is factorable and becomes a product of two orthogonal lines (see the right of Fig.\ref{twoline}).
	Otherwise, 
	\begin{equation}
		y^2 + \frac{2 a}{b} y x - x^2 - \frac{a^2 + b^2}{b} y + 1 = 0 \label{d2normal}
	\end{equation}
	If Eq.~\eqref{d2normal} is irreducible, 
	it is an orthogonal hyperbola (see Fig.\ref{hyperbola}).
	If Eq.~\eqref{d2normal} is factorable, 
	\begin{equation}
		(c_1 y + c_2 x + c_3) ( \frac{1}{c_1} y - \frac{1}{c_2} x + \frac{1}{c_3} ) = 0 \label{p9}
	\end{equation}
	These two lines are orthogonal to each other as in the left of Fig.\ref{twoline}.
\end{proof}
\begin{theorem}
\label{th:d2f}
	An Equation of an edge of angular Voronoi diagram is exactly degree two and factorable, 
	if and only if it is a product of two orthogonal lines equation.
\end{theorem}
\begin{proof}
	See the proof of Th.~\ref{th:d2u}.
\end{proof}
\begin{theorem}
\label{th:d1}
	An equation of an edge of an angular Voronoi diagram is never degree one.
\end{theorem}
\begin{proof}
	Eliminating the degree two terms from Eq.~\eqref{d2eq}, 
	we obtain $ a = b = 0 $.
	This means the two line segments are exactly the same.
	This case is meaningless.
\end{proof}

\section{Conclusions}
\label{conclusion}
\begin{table}
	\caption{The types of angular Voronoi diagram's edges}
	\label{typesAVD}
	\begin{center}
		\begin{tabular}{|l|l|l|}
			\hline
			degree 3	&	irreducible					&	no singularity								\\
						&								&	(most general case)							\\ \cline{3-3}
						&								&	one singularity(Ex.\ref{th:d3u})			\\ \cline{2-3}
						&	\multicolumn{2}{|l|}{factorable(circle $ \times $ line)(Th.\ref{th:d3f})}	\\ \hline
			degree 2	&	\multicolumn{2}{|l|}{irreducible(hyperbola)(Th.\ref{th:d2u})}				\\ \cline{2-3}
						&	\multicolumn{2}{|l|}{factorable(two line)(Th.\ref{th:d2f})}					\\ \hline
			degree 1	&	\multicolumn{2}{|l|}{unrealizable(Th.\ref{th:d1})}							\\ \hline
		\end{tabular}
	\end{center}
\end{table}

We classified possible curves that can appear as an edge of an angular
Voronoi diagram. Although we have not got a necessary and sufficient
condition for each case, we successfully enumerated the cases when
the edge becomes somewhat ``unusual.''

The summary of our results is shown in Table \ref{typesAVD}.
We showed that a Voronoi edge can be of degree three or two, but cannot
be of degree one. A degree three edge can have a singularity and
if it can be factored into two curves, they are always a circle and a
line. A degree two curve is a hyperbola when it is irreducible, 
and it is two lines when it is factorable. 

Now we are interested in a further analysis on conditions for unusual
curves. For example, ``What is the necessary and sufficient condition
for an edge to be a circle and a line?'' is a problem to
consider. Additionally, a possible relative position of multiple edges is
also important. To check if what kind of bad-conditioned position (for
example, the case multiple curves are crossing at their singular point)
can be possible is essential to achieve computational robustness.

\bibliographystyle{abbrv}
\bibliography{isvd07muta}

\begin{thebibliography}{1}

\bibitem{avd}
T.~Asano, H.~Tamaki, N.~Katoh, and T.~Tokuyama.
\newblock Angular voronoi diagram with applications.
\newblock In {\em 3rd International Symposium on Voronoi Diagrams in Science
  and Engineering (ISVD'06)}, pages 18--24, 2006.

\bibitem{vdsurvey}
F.~Aurenhammer.
\newblock Voronoi diagrams - a survey of a fundamental geometric data
  structure.
\newblock {\em ACM Comput. Surv.}, 23(3):345--405, 1991.

\bibitem{Hartshorn77}
R.~Hartshorn.
\newblock {\em Algebraic Geometry}.
\newblock Springer--Verlag, 1977.

\bibitem{kato05}
K.~Kato, M.~Oto, H.~Imai, and K.~Imai.
\newblock Voronoi diagrams for pure 1-qubit quantum states.
\newblock In {\em Proceedings of International Symposium on Voronoi Diagram},
  pages 293--299, Seoul, Korea, 2005.

\bibitem{kato06a}
K.~Kato, M.~Oto, H.~Imai, and K.~Imai.
\newblock On a geometric structure of pure multi-qubit quantum states and its
  applicability to a numerical computation.
\newblock In {\em Proceedings of International Symposium on Voronoi Diagram},
  pages 48--53, Banff, Canada, 2006.

\bibitem{kato06b}
K.~Kato, M.~Oto, H.~Imai, and K.~Imai.
\newblock Voronoi diagrams and a numerical estimation of a quantum channel
  capacity.
\newblock In {\em 2nd Doctoral Workshop on Mathematical and Engineering Methods
  in Computer Science (MEMICS 2006)}, pages 69--76, Mikulov, Czech, Oct. 2006.

\bibitem{spatialTessellations}
A.~Okabe, B.~Boots, K.~Sugihara, and S.~N. Chiu.
\newblock {\em Spatial Tessellations, Concepts and Applications of Voronoi
  Diagrams}.
\newblock John Wiley \& Sons, New York, 2nd edition, 2000.

\end{thebibliography}

\end{document}